\documentstyle[epsfig,12pt]{article}

\begin{document}

\begin{titlepage}

\pagestyle{empty}

\vskip 2.0 cm

\begin{center}
{\large EUROPEAN LABORATORY FOR PARTICLE PHYSICS (CERN)}
\end{center}

\vskip 1.0 cm

\begin{flushright}
CERN-EP/98-101 \\
17 June 1998\\
\end{flushright}

\vspace{1. cm}

\begin{center}
\boldmath
{\Large\bf   The Forward-Backward Asymmetry for Charm Quarks
             at the Z }
\unboldmath

\vskip .8 cm

{\bf The ALEPH Collaboration\footnote{See next pages for the list
    of authors}}\\
\end{center}

\vskip  2cm

\begin{center}
{\bf Abstract}
\end{center}
The data set collected with the ALEPH detector from 1991 to 1995 
at {\sc LEP} has been analysed to measure the
charm forward-backward asymmetry at the Z. 
Out of a total of 4.1 million hadronic Z decays, about 36000 
high momentum $\mathrm{D^{*+}}$, $\mathrm{D^{+}}$ and
$\mathrm{D^{0}}$ decays were  
reconstructed, of which 80\% originate from  
$\mathrm{Z \to c\bar{c}}$ events. The forward-backward asymmetry was
  measured at three 
energy points:    
$$
\begin{array}{rcl}
A_{FB}^{\mathrm{c}}(\sqrt{s} = 89.37~\mathrm{GeV})&=&\left( -1.0\pm 4.4\right)\!\%  \\
A_{FB}^{\mathrm{c}}(\sqrt{s} = 91.22~\mathrm{GeV})&=&\left( 6.3\pm 1.0\right)\!\%  \\
A_{FB}^{\mathrm{c}}(\sqrt{s} = 92.96~\mathrm{GeV})&=&\left( 11.0\pm 3.4\right)\!\% \ . 
\end{array}
$$
From this analysis, a value of the effective electroweak mixing angle 
$\sin^2 \theta^{\mathrm eff}_W = 0.2321 \pm 0.0016$ is extracted.

\vskip 2cm

\begin{center}
{\it (To be submitted to Physics Letters B)}
\end{center}

\end{titlepage}

\pagestyle{empty}
\newpage
\small
%
\newlength{\saveparskip}
\newlength{\savetextheight}
\newlength{\savetopmargin}
\newlength{\savetextwidth}
\newlength{\saveoddsidemargin}
\newlength{\savetopsep}
\setlength{\saveparskip}{\parskip}
\setlength{\savetextheight}{\textheight}
\setlength{\savetopmargin}{\topmargin}
\setlength{\savetextwidth}{\textwidth}
\setlength{\saveoddsidemargin}{\oddsidemargin}
\setlength{\savetopsep}{\topsep}
%
%
\setlength{\parskip}{0.0cm}
\setlength{\textheight}{25.0cm}
\setlength{\topmargin}{-1.5cm}
\setlength{\textwidth}{16 cm}
\setlength{\oddsidemargin}{-0.0cm}
\setlength{\topsep}{1mm}
\pretolerance=10000
\centerline{\large\bf The ALEPH Collaboration}
\footnotesize
\vspace{0.5cm}
{\raggedbottom
\begin{sloppypar}
\samepage\noindent
R.~Barate,
D.~Buskulic,
D.~Decamp,
P.~Ghez,
C.~Goy,
\mbox{J.-P.~Lees},
A.~Lucotte,
E.~Merle,
\mbox{M.-N.~Minard},
\mbox{J.-Y.~Nief},
B.~Pietrzyk
\nopagebreak
\begin{center}
\parbox{15.5cm}{\sl\samepage
Laboratoire de Physique des Particules (LAPP), IN$^{2}$P$^{3}$-CNRS,
F-74019 Annecy-le-Vieux Cedex, France}
\end{center}\end{sloppypar}
\vspace{2mm}
\begin{sloppypar}
\noindent
R.~Alemany,
G.~Boix,
M.P.~Casado,
M.~Chmeissani,
J.M.~Crespo,
M.~Delfino,
E.~Fernandez,
\mbox{M.~Fernandez-Bosman},
Ll.~Garrido,$^{15}$
E.~Graug\`{e}s,
A.~Juste,
M.~Martinez,
G.~Merino,
R.~Miquel,
Ll.M.~Mir,
I.C.~Park,
A.~Pascual,
I.~Riu,
F.~Sanchez
\nopagebreak
\begin{center}
\parbox{15.5cm}{\sl\samepage
Institut de F\'{i}sica d'Altes Energies, Universitat Aut\`{o}noma
de Barcelona, E-08193 Bellaterra (Barcelona), Spain$^{7}$}
\end{center}\end{sloppypar}
\vspace{2mm}
\begin{sloppypar}
\noindent
A.~Colaleo,
D.~Creanza,
M.~de~Palma,
G.~Gelao,
G.~Iaselli,
G.~Maggi,
M.~Maggi,
S.~Nuzzo,
A.~Ranieri,
G.~Raso,
F.~Ruggieri,
G.~Selvaggi,
L.~Silvestris,
P.~Tempesta,
A.~Tricomi,$^{3}$
G.~Zito
\nopagebreak
\begin{center}
\parbox{15.5cm}{\sl\samepage
Dipartimento di Fisica, INFN Sezione di Bari, I-70126
Bari, Italy}
\end{center}\end{sloppypar}
\vspace{2mm}
\begin{sloppypar}
\noindent
X.~Huang,
J.~Lin,
Q. Ouyang,
T.~Wang,
Y.~Xie,
R.~Xu,
S.~Xue,
J.~Zhang,
L.~Zhang,
W.~Zhao
\nopagebreak
\begin{center}
\parbox{15.5cm}{\sl\samepage
Institute of High-Energy Physics, Academia Sinica, Beijing, The People's
Republic of China$^{8}$}
\end{center}\end{sloppypar}
\vspace{2mm}
\begin{sloppypar}
\noindent
D.~Abbaneo,
U.~Becker,
\mbox{P.~Bright-Thomas},$^{24}$
D.~Casper,
M.~Cattaneo,
F.~Cerutti,
V.~Ciulli,
G.~Dissertori,
H.~Drevermann,
R.W.~Forty,
M.~Frank,
R.~Hagelberg,
A.W. Halley,
J.B.~Hansen,
J.~Harvey,
P.~Janot,
B.~Jost,
I.~Lehraus,
P.~Mato,
A.~Minten,
L.~Moneta,$^{21}$
A.~Pacheco,
F.~Ranjard,
L.~Rolandi,
D.~Rousseau,
D.~Schlatter,
M.~Schmitt,$^{20}$
O.~Schneider,
W.~Tejessy,
F.~Teubert,
I.R.~Tomalin,
H.~Wachsmuth
\nopagebreak
\begin{center}
\parbox{15.5cm}{\sl\samepage
European Laboratory for Particle Physics (CERN), CH-1211 Geneva 23,
Switzerland}
\end{center}\end{sloppypar}
\vspace{2mm}
\begin{sloppypar}
\noindent
Z.~Ajaltouni,
F.~Badaud,
G.~Chazelle,
O.~Deschamps,
A.~Falvard,
C.~Ferdi,
P.~Gay,
C.~Guicheney,
P.~Henrard,
J.~Jousset,
B.~Michel,
S.~Monteil,
\mbox{J-C.~Montret},
D.~Pallin,
P.~Perret,
F.~Podlyski,
J.~Proriol,
P.~Rosnet
\nopagebreak
\begin{center}
\parbox{15.5cm}{\sl\samepage
Laboratoire de Physique Corpusculaire, Universit\'e Blaise Pascal,
IN$^{2}$P$^{3}$-CNRS, Clermont-Ferrand, F-63177 Aubi\`{e}re, France}
\end{center}\end{sloppypar}
\vspace{2mm}
\begin{sloppypar}
\noindent
J.D.~Hansen,
J.R.~Hansen,
P.H.~Hansen,
B.S.~Nilsson,
B.~Rensch,
A.~W\"a\"an\"anen
\begin{center}
\parbox{15.5cm}{\sl\samepage
Niels Bohr Institute, DK-2100 Copenhagen, Denmark$^{9}$}
\end{center}\end{sloppypar}
\vspace{2mm}
\begin{sloppypar}
\noindent
G.~Daskalakis,
A.~Kyriakis,
C.~Markou,
E.~Simopoulou,
I.~Siotis,
A.~Vayaki
\nopagebreak
\begin{center}
\parbox{15.5cm}{\sl\samepage
Nuclear Research Center Demokritos (NRCD), GR-15310 Attiki, Greece}
\end{center}\end{sloppypar}
\vspace{2mm}
\begin{sloppypar}
\noindent
A.~Blondel,
G.~Bonneaud,
\mbox{J.-C.~Brient},
P.~Bourdon,
A.~Roug\'{e},
M.~Rumpf,
A.~Valassi,$^{6}$
M.~Verderi,
H.~Videau
\nopagebreak
\begin{center}
\parbox{15.5cm}{\sl\samepage
Laboratoire de Physique Nucl\'eaire et des Hautes Energies, Ecole
Polytechnique, IN$^{2}$P$^{3}$-CNRS, \mbox{F-91128} Palaiseau Cedex, France}
\end{center}\end{sloppypar}
\vspace{2mm}
\begin{sloppypar}
\noindent
E.~Focardi,
G.~Parrini,
K.~Zachariadou
\nopagebreak
\begin{center}
\parbox{15.5cm}{\sl\samepage
Dipartimento di Fisica, Universit\`a di Firenze, INFN Sezione di Firenze,
I-50125 Firenze, Italy}
\end{center}\end{sloppypar}
\vspace{2mm}
\begin{sloppypar}
\noindent
M.~Corden,
C.~Georgiopoulos,
D.E.~Jaffe
\nopagebreak
\begin{center}
\parbox{15.5cm}{\sl\samepage
Supercomputer Computations Research Institute,
Florida State University,
Tallahassee, FL 32306-4052, USA $^{13,14}$}
\end{center}\end{sloppypar}
\vspace{2mm}
\begin{sloppypar}
\noindent
A.~Antonelli,
G.~Bencivenni,
G.~Bologna,$^{4}$
F.~Bossi,
P.~Campana,
G.~Capon,
V.~Chiarella,
G.~Felici,
P.~Laurelli,
G.~Mannocchi,$^{5}$
F.~Murtas,
G.P.~Murtas,
L.~Passalacqua,
\mbox{M.~Pepe-Altarelli}
\nopagebreak
\begin{center}
\parbox{15.5cm}{\sl\samepage
Laboratori Nazionali dell'INFN (LNF-INFN), I-00044 Frascati, Italy}
\end{center}\end{sloppypar}
\vspace{2mm}
\begin{sloppypar}
\noindent
L.~Curtis,
J.G.~Lynch,
P.~Negus,
V.~O'Shea,
C.~Raine,
J.M.~Scarr,
K.~Smith,
\mbox{P.~Teixeira-Dias},
A.S.~Thompson,
E.~Thomson
\nopagebreak
\begin{center}
\parbox{15.5cm}{\sl\samepage
Department of Physics and Astronomy, University of Glasgow, Glasgow G12
8QQ,United Kingdom$^{10}$}
\end{center}\end{sloppypar}
\pagebreak
\vspace{2mm}
\begin{sloppypar}
\noindent
O.~Buchm\"uller,
S.~Dhamotharan,
C.~Geweniger,
G.~Graefe,
P.~Hanke,
G.~Hansper,
V.~Hepp,
E.E.~Kluge,
A.~Putzer,
J.~Sommer,
K.~Tittel,
S.~Werner,
M.~Wunsch
\nopagebreak
\begin{center}
\parbox{15.5cm}{\sl\samepage
Institut f\"ur Hochenergiephysik, Universit\"at Heidelberg, D-69120
Heidelberg, Germany$^{16}$}
\end{center}\end{sloppypar}
\vspace{2mm}
\begin{sloppypar}
\noindent
R.~Beuselinck,
D.M.~Binnie,
W.~Cameron,
P.J.~Dornan,$^{2}$
M.~Girone,
S.~Goodsir,
E.B.~Martin,
N.~Marinelli,
A.~Moutoussi,
J.~Nash,
J.K.~Sedgbeer,
P.~Spagnolo,
M.D.~Williams
\nopagebreak
\begin{center}
\parbox{15.5cm}{\sl\samepage
Department of Physics, Imperial College, London SW7 2BZ,
United Kingdom$^{10}$}
\end{center}\end{sloppypar}
\vspace{2mm}
\begin{sloppypar}
\noindent
V.M.~Ghete,
P.~Girtler,
E.~Kneringer,
D.~Kuhn,
G.~Rudolph
\nopagebreak
\begin{center}
\parbox{15.5cm}{\sl\samepage
Institut f\"ur Experimentalphysik, Universit\"at Innsbruck, A-6020
Innsbruck, Austria$^{18}$}
\end{center}\end{sloppypar}
\vspace{2mm}
\begin{sloppypar}
\noindent
A.P.~Betteridge,
C.K.~Bowdery,
P.G.~Buck,
P.~Colrain,
G.~Crawford,
A.J.~Finch,
F.~Foster,
G.~Hughes,
R.W.L.~Jones,
N.A.~Robertson,
M.I.~Williams
\nopagebreak
\begin{center}
\parbox{15.5cm}{\sl\samepage
Department of Physics, University of Lancaster, Lancaster LA1 4YB,
United Kingdom$^{10}$}
\end{center}\end{sloppypar}
\vspace{2mm}
\begin{sloppypar}
\noindent
I.~Giehl,
C.~Hoffmann,
K.~Jakobs,
K.~Kleinknecht,
G.~Quast,
B.~Renk,
E.~Rohne,
\mbox{H.-G.~Sander},
P.~van~Gemmeren,
C.~Zeitnitz
\nopagebreak
\begin{center}
\parbox{15.5cm}{\sl\samepage
Institut f\"ur Physik, Universit\"at Mainz, D-55099 Mainz, Germany$^{16}$}
\end{center}\end{sloppypar}
\vspace{2mm}
\begin{sloppypar}
\noindent
J.J.~Aubert,
C.~Benchouk,
A.~Bonissent,
G.~Bujosa,
J.~Carr,$^{2}$
P.~Coyle,
F.~Etienne,
O.~Leroy,
F.~Motsch,
P.~Payre,
M.~Talby,
A.~Sadouki,
M.~Thulasidas,
K.~Trabelsi
\nopagebreak
\begin{center}
\parbox{15.5cm}{\sl\samepage
Centre de Physique des Particules, Facult\'e des Sciences de Luminy,
IN$^{2}$P$^{3}$-CNRS, F-13288 Marseille, France}
\end{center}\end{sloppypar}
\vspace{2mm}
\begin{sloppypar}
\noindent
M.~Aleppo,
M.~Antonelli,
F.~Ragusa
\nopagebreak
\begin{center}
\parbox{15.5cm}{\sl\samepage
Dipartimento di Fisica, Universit\`a di Milano e INFN Sezione di Milano,
I-20133 Milano, Italy}
\end{center}\end{sloppypar}
\vspace{2mm}
\begin{sloppypar}
\noindent
R.~Berlich,
V.~B\"uscher,
G.~Cowan,
H.~Dietl,
G.~Ganis,
G.~L\"utjens,
C.~Mannert,
W.~M\"anner,
\mbox{H.-G.~Moser},
S.~Schael,
R.~Settles,
H.~Seywerd,
H.~Stenzel,
W.~Wiedenmann,
G.~Wolf
\nopagebreak
\begin{center}
\parbox{15.5cm}{\sl\samepage
Max-Planck-Institut f\"ur Physik, Werner-Heisenberg-Institut,
D-80805 M\"unchen, Germany\footnotemark[16]}
\end{center}\end{sloppypar}
\vspace{2mm}
\begin{sloppypar}
\noindent
J.~Boucrot,
O.~Callot,
S.~Chen,
A.~Cordier,
M.~Davier,
L.~Duflot,
\mbox{J.-F.~Grivaz},
Ph.~Heusse,
A.~H\"ocker,
A.~Jacholkowska,
D.W.~Kim,$^{12}$
F.~Le~Diberder,
J.~Lefran\c{c}ois,
\mbox{A.-M.~Lutz},
\mbox{M.-H.~Schune},
E.~Tournefier,
\mbox{J.-J.~Veillet},
I.~Videau,
D.~Zerwas
\nopagebreak
\begin{center}
\parbox{15.5cm}{\sl\samepage
Laboratoire de l'Acc\'el\'erateur Lin\'eaire, Universit\'e de Paris-Sud,
IN$^{2}$P$^{3}$-CNRS, F-91898 Orsay Cedex, France}
\end{center}\end{sloppypar}
\vspace{2mm}
\begin{sloppypar}
\noindent
\samepage
P.~Azzurri,
G.~Bagliesi,$^{2}$
G.~Batignani,
S.~Bettarini,
T.~Boccali,
C.~Bozzi,
G.~Calderini,
M.~Carpinelli,
M.A.~Ciocci,
R.~Dell'Orso,
R.~Fantechi,
I.~Ferrante,
L.~Fo\`{a},$^{1}$
F.~Forti,
A.~Giassi,
M.A.~Giorgi,
A.~Gregorio,
F.~Ligabue,
A.~Lusiani,
P.S.~Marrocchesi,
A.~Messineo,
F.~Palla,
G.~Rizzo,
G.~Sanguinetti,
A.~Sciab\`a,
G.~Sguazzoni,
R.~Tenchini,
G.~Tonelli,$^{19}$
C.~Vannini,
A.~Venturi,
P.G.~Verdini
\samepage
\begin{center}
\parbox{15.5cm}{\sl\samepage
Dipartimento di Fisica dell'Universit\`a, INFN Sezione di Pisa,
e Scuola Normale Superiore, I-56010 Pisa, Italy}
\end{center}\end{sloppypar}
\vspace{2mm}
\begin{sloppypar}
\noindent
G.A.~Blair,
L.M.~Bryant,
J.T.~Chambers,
M.G.~Green,
T.~Medcalf,
P.~Perrodo,
J.A.~Strong,
\mbox{J.H.~von~Wimmersperg-Toeller}
\nopagebreak
\begin{center}
\parbox{15.5cm}{\sl\samepage
Department of Physics, Royal Holloway \& Bedford New College,
University of London, Surrey TW20 OEX, United Kingdom$^{10}$}
\end{center}\end{sloppypar}
\vspace{2mm}
\begin{sloppypar}
\noindent
D.R.~Botterill,
R.W.~Clifft,
T.R.~Edgecock,
P.R.~Norton,
J.C.~Thompson,
A.E.~Wright
\nopagebreak
\begin{center}
\parbox{15.5cm}{\sl\samepage
Particle Physics Dept., Rutherford Appleton Laboratory,
Chilton, Didcot, Oxon OX11 OQX, United Kingdom$^{10}$}
\end{center}\end{sloppypar}
\vspace{2mm}
\begin{sloppypar}
\noindent
\mbox{B.~Bloch-Devaux},
P.~Colas,
S.~Emery,
W.~Kozanecki,
E.~Lan\c{c}on,$^{2}$
\mbox{M.-C.~Lemaire},
E.~Locci,
P.~Perez,
J.~Rander,
\mbox{J.-F.~Renardy},
A.~Roussarie,
\mbox{J.-P.~Schuller},
J.~Schwindling,
A.~Trabelsi,
B.~Vallage
\nopagebreak
\begin{center}
\parbox{15.5cm}{\sl\samepage
CEA, DAPNIA/Service de Physique des Particules,
CE-Saclay, F-91191 Gif-sur-Yvette Cedex, France$^{17}$}
\end{center}\end{sloppypar}
\pagebreak
\vspace{2mm}
\begin{sloppypar}
\noindent
S.N.~Black,
J.H.~Dann,
R.P.~Johnson,
H.Y.~Kim,
N.~Konstantinidis,
A.M.~Litke,
M.A. McNeil,
G.~Taylor
\nopagebreak
\begin{center}
\parbox{15.5cm}{\sl\samepage
Institute for Particle Physics, University of California at
Santa Cruz, Santa Cruz, CA 95064, USA$^{22}$}
\end{center}\end{sloppypar}
\vspace{2mm}
\begin{sloppypar}
\noindent
C.N.~Booth,
S.~Cartwright,
F.~Combley,
M.S.~Kelly,
M.~Lehto,
L.F.~Thompson
\nopagebreak
\begin{center}
\parbox{15.5cm}{\sl\samepage
Department of Physics, University of Sheffield, Sheffield S3 7RH,
United Kingdom$^{10}$}
\end{center}\end{sloppypar}
\vspace{2mm}
\begin{sloppypar}
\noindent
K.~Affholderbach,
A.~B\"ohrer,
S.~Brandt,
C.~Grupen,
P.~Saraiva,
L.~Smolik,
F.~Stephan
\nopagebreak
\begin{center}
\parbox{15.5cm}{\sl\samepage
Fachbereich Physik, Universit\"at Siegen, D-57068 Siegen,
 Germany$^{16}$}
\end{center}\end{sloppypar}
\vspace{2mm}
\begin{sloppypar}
\noindent
G.~Giannini,
B.~Gobbo,
G.~Musolino
\nopagebreak
\begin{center}
\parbox{15.5cm}{\sl\samepage
Dipartimento di Fisica, Universit\`a di Trieste e INFN Sezione di Trieste,
I-34127 Trieste, Italy}
\end{center}\end{sloppypar}
\vspace{2mm}
\begin{sloppypar}
\noindent
J.~Rothberg,
S.~Wasserbaech
\nopagebreak
\begin{center}
\parbox{15.5cm}{\sl\samepage
Experimental Elementary Particle Physics, University of Washington, WA 98195
Seattle, U.S.A.}
\end{center}\end{sloppypar}
\vspace{2mm}
\begin{sloppypar}
\noindent
S.R.~Armstrong,
E.~Charles,
P.~Elmer,
D.P.S.~Ferguson,
Y.~Gao,
S.~Gonz\'{a}lez,
T.C.~Greening,
O.J.~Hayes,
H.~Hu,
S.~Jin,
P.A.~McNamara III,
J.M.~Nachtman,$^{23}$
J.~Nielsen,
W.~Orejudos,
Y.B.~Pan,
Y.~Saadi,
I.J.~Scott,
J.~Walsh,
Sau~Lan~Wu,
X.~Wu,
G.~Zobernig
\nopagebreak
\begin{center}
\parbox{15.5cm}{\sl\samepage
Department of Physics, University of Wisconsin, Madison, WI 53706,
USA$^{11}$}
\end{center}\end{sloppypar}
}
\footnotetext[1]{Now at CERN, 1211 Geneva 23,
Switzerland.}
\footnotetext[2]{Also at CERN, 1211 Geneva 23, Switzerland.}
\footnotetext[3]{Also at Dipartimento di Fisica, INFN, Sezione di Catania, Catania, Italy.}
\footnotetext[4]{Also Istituto di Fisica Generale, Universit\`{a} di
Torino, Torino, Italy.}
\footnotetext[5]{Also Istituto di Cosmo-Geofisica del C.N.R., Torino,
Italy.}
\footnotetext[6]{Supported by the Commission of the European Communities,
contract ERBCHBICT941234.}
\footnotetext[7]{Supported by CICYT, Spain.}
\footnotetext[8]{Supported by the National Science Foundation of China.}
\footnotetext[9]{Supported by the Danish Natural Science Research Council.}
\footnotetext[10]{Supported by the UK Particle Physics and Astronomy Research
Council.}
\footnotetext[11]{Supported by the US Department of Energy, grant
DE-FG0295-ER40896.}
\footnotetext[12]{Permanent address: Kangnung National University, Kangnung, 
Korea.}
\footnotetext[13]{Supported by the US Department of Energy, contract
DE-FG05-92ER40742.}
\footnotetext[14]{Supported by the US Department of Energy, contract
DE-FC05-85ER250000.}
\footnotetext[15]{Permanent address: Universitat de Barcelona, 08208 Barcelona,
Spain.}
\footnotetext[16]{Supported by the Bundesministerium f\"ur Bildung,
Wissenschaft, Forschung und Technologie, Germany.}
\footnotetext[17]{Supported by the Direction des Sciences de la
Mati\`ere, C.E.A.}
\footnotetext[18]{Supported by Fonds zur F\"orderung der wissenschaftlichen
Forschung, Austria.}
\footnotetext[19]{Also at Istituto di Matematica e Fisica,
Universit\`a di Sassari, Sassari, Italy.}
\footnotetext[20]{Now at Harvard University, Cambridge, MA 02138, U.S.A.}
\footnotetext[21]{Now at University of Geneva, 1211 Geneva 4, Switzerland.}
\footnotetext[22]{Supported by the US Department of Energy,
grant DE-FG03-92ER40689.}
\footnotetext[23]{Now at University of California at Los Angeles (UCLA),
Los Angeles, CA 90024, U.S.A.}
\footnotetext[24]{Now at School of Physics and Astronomy,
Birmingham B15 2TT, U.K.}
%
%
\setlength{\parskip}{\saveparskip}
\setlength{\textheight}{\savetextheight}
\setlength{\topmargin}{\savetopmargin}
\setlength{\textwidth}{\savetextwidth}
\setlength{\oddsidemargin}{\saveoddsidemargin}
\setlength{\topsep}{\savetopsep}
\normalsize
\newpage
\pagestyle{plain}
\setcounter{page}{1}


%
\def \etal  {\hbox{\it et al.}}
\def \zphys {Z. Phys.~}
\def \NIM   {Nucl. Instr. Methods~}
\def \PL    {Phys. Lett.~}
\def \epj   {Eur. Phys. J. C}
%
%
\def \LEP    {{\sc LEP}}
\def \A      {{ALEPH}}
\def \Ab     {{{ALEPH}}}
\def \TPC    {{\sc tpc}}
\def \CLEO   {{\sc Cleo}}
\def \DELC   {{\sc Delco}}
\def \MARK   {{\sc Mark~III}}
\def \ARGUS  {{\sc Argus}}

\def \dedx   {{dE/dx}}
\def \ia            {{\it a}}
\def \iv            {{\it v}}
\def \u             {\mathrm u}
\def \b             {\mathrm b}
\def \c             {\mathrm c}
\def \d             {\mathrm d}
\def \e             {\mathrm e}
\def \q             {\mathrm q}
\def \s             {\mathrm s}
\def \f             {\mathrm f}
\def \qq            {\q \bar{q}}
\def \thr           {{\mathrm{thrust}}}

\def \D             {\mathrm D}
\def \dstp           {\D^{*+}}
\def \K             {\mathrm K}
\def \Z             {\mathrm Z}
\def \Zcc           {\Z \to \c \bar{\c} }
\def \Zbb           {\Z \to \b \bar{\b} }
\def \Zff           {\Z \to \f \bar{\f} }
\def \Zuu           {\Z \to \u \bar{\u} }
\def \decDS         {\D^{*+} \to \D^0 \pi^+}
\def \GeV           {{\mathrm GeV}}
\def \GeVc          {\GeV/c}
\def \GeVcc         {\GeV/c^2}
\def \MeV           {{\mathrm MeV}}
\def \MeVc          {\MeV/c}
\def \MeVcc         {\MeV/c^2}
\def \deckp         {\D^{0} \to \K^- \pi^+}
\def \dpkpp         {\D^{+} \to \K^- \pi^+ \pi^+}
\def \deckppp       {\D^{0} \to \K^- \pi^+ \pi^+ \pi^-}
\def \deckpp        {\D^{0} \to \K^- \pi^+ \pi^0}
\def \deckpS        {\D^{0} \to \K^- \pi^+ (\pi^0)}
\def \decdst        {\D^{*+} \to \pi^{+}_{s} \D^0}
\def \decskp        {\D^{*+} \to \pi^{+}_{s} \K^- \pi^+}
\def \decskppp      {\D^{*+} \to \pi^{+}_{s} \K^- \pi^+ \pi^+ \pi^-}
\def \decskpp       {\D^{*+} \to \pi^{+}_{s} \K^- \pi^+ \pi^0}
\def \decskpS       {\D^{*+} \to \pi^{+}_{s} \K^- \pi^+ (\pi^0)}
\def \epsc          {\varepsilon_{\c}}
\def \epsb          {\varepsilon_{\b}}
\def \pctod         {P_{\c \to \D^*}}
\def \pbtod         {P_{\b \to \D^*}}

\def \rightdownarrow
 {\kern.5em
 \rule[.55ex]{.15mm}{2.5mm}
 {\mbox{$\kern0.em{\rightarrow}$}}}

\newcommand{\afb}[1]{A_{FB}^{#1}}

\pagestyle{plain}
\setcounter{page}{1}
\setcounter{footnote}{0}

\section{Introduction}

The forward-backward asymmetry in $\Zcc$ decays
provides a direct and precise test of the coupling of the Z to 
up-type quarks. The asymmetry $\afb{\f}$ in $\Zff$ decays arises from 
parity violation in Z production and decay. In the Standard Model, the
differential 
cross section, expressed as a function of the angle $\theta$ between
the outgoing fermion and the incoming electron, is
\begin{equation}
\frac{1}{\sigma} {d \sigma \over d \cos \theta } = 
\frac{3}{8} (1 + \cos^2 \theta) + \afb{\f} \cos \theta \ 
\label{eq:dxsec}
\end{equation}
At the $\rm{Z}$ pole, for unpolarised $e^+ e^-$ beams, $\afb{\f}$  
is related to the pole asymmetry $\afb{0,\f}$, defined in terms
of the effective couplings in the improved Born approximation as 
\begin{equation}
{A_{FB}^{0,\f}} = {{3}\over{4}}
 {{2 \mathrm{ g_{Ve} g_{Ae}}} \over {\mathrm{( {g^2_{Ve}} + {g^2_{Ae}}})}}
 {{2 \mathrm{ g_{Vf} g_{Af}}} \over {\mathrm{( {g^2_{Vf}} + {g^2_{Af}}})}}\ .
\label{eq:pole}
\end{equation}
The measurement can be interpreted in terms of the effective 
electroweak mixing angle $ \sin^2 \theta_W^{\mathrm eff} = {{1} \over {4}}
(1- {\mathrm{{g_{Ve}}}}/{\mathrm{{g_{Ae}}}})$.

The $\Zcc$ decays provide a convenient way to directly test
the Z coupling to up-type quarks, since the $\Zuu$ events are
much harder to isolate. In this paper a sample of $\Zcc$  decays is selected
using high energy $\D^{+}$, $\D^{0}$ and
$\D^{*+}$, fully reconstructed from their decays products. The 
restricted number of usable decay channels limits the tagging
efficiency. On the other hand, the high purity obtained in the  
selection, together with a minor dependence on the decay models, allow
a measurement with small systematic uncertainties. 

First the selection of charmed mesons is described,
then the background estimate is discussed and the asymmetry is
measured on the selected sample at three different centre of mass energies: at
the Z peak and at $\pm2~\GeV$ off-peak.

\boldmath
\section{Reconstruction of $\D^{(*)}$\/ mesons}
\unboldmath

A detailed description of the \A\ detector and its performance can be
found in Ref.~\cite{nim,perf}. Charged particles are detected in the
central part of the detector, consisting of a two-layer silicon vertex
detector with double-sided ($r$-$\phi$ and $z$)   
readout, a cylindrical drift chamber and a large time projection
chamber (\TPC), which together measure up to 33 coordinates along the
charged particle trajectories.
Tracking is performed in a 1.5~T axial magnetic field provided by a
superconducting solenoid.
The \TPC\ also provides up to 338 measurements of ionization 
($\dedx$) allowing particle identification.
The electromagnetic calorimeter is a
lead/wire-chamber sandwich operated in proportional mode. It is read
out in projective towers of typically $15 \times 15$~mrad$^2$ size
segmented in three longitudinal sections. The iron
return yoke is instrumented with streamer tubes to provide a
measurement of the hadronic energy. 
An energy flow algorithm~\cite{perf} combines charged particles
momenta and calorimetric energy measurements and provides a list of
energy flow particles on which the analysis is based. 

About four million hadronic Z decays 
are selected as described in Ref.~\cite{SELE}, 
out of the data set collected by \A\ during the 1991--1995
running period at the Z resonance.

Charmed mesons are reconstructed in the following decay modes
(charge-conjugate modes are implied throughout the paper): 
$$
\begin{array}{l@{\ \ \ }ll@{\hspace{1.cm}}l@{\ \ \ }ll}
({i})   & \D^{*+} \to & \D^0 \pi^+_s     & 
({ii})  & \D^{*+} \to & \D^0 \pi^+_s     \\ 
        &             & \rightdownarrow \K^-\pi^+  &
        &             & \rightdownarrow \K^-\pi^+\pi^0  \\[0.3em]
({iii}) & \D^0    \to & \K^-\pi^+      & 
({iv})  & \D^{*+} \to & \D^0 \pi^+_s     \\ 
        &             &                &
        &             & \rightdownarrow \K^-\pi^+\pi^+\pi^-  \\[0.3em]
({v})   & \D^{*+} \to & \D^0 \pi^+_s     & 
({vi})  & \D^+    \to & \K^-\pi^+\pi^+ \\
        &             & \rightdownarrow \K^-\pi^+ +\ (\pi^0)\  
        &             &                & 
\end{array}
$$
where the soft pion from $\D^{*+}$\/ decay is indicated as $\pi^+_s$. 
Channel ({\it v}) is 
selected without reconstructing the $\pi^0$\/ by using 
the kinematic properties of the underlying  resonances in the $\D^0$\/ decay
that make the $\K^{-} \pi^+$\/ invariant mass peak near $1.6~\GeVcc$. 
The reconstruction proceeds from channel ({\it i}) to channel ({\it vi});
once a candidate has been found in one event, this event is excluded 
in the following channels. Consequently, the $\deckp$ decays from
  $\D^{*+}$ are removed from the inclusive sample (${\it iii}$), thus avoiding
 double counting. The number of candidates for each selected sample is
 listed in Table~\ref{tabsel}. 

\begin{table}[t]
\caption[]{\it Results of the  $\D^{+}$, $\D^{0}$ and $\D^{*+}$
  reconstruction. The second column shows 
  the number of candidates in the different channels, while the last column
  shows the number of signal events, after combinatorial
  background subtraction, together with the statistical
  and systematic uncertainties.}
\begin{center}
\begin{tabular}{|l|r|r@{$\pm$}r@{$\pm$}r|}
\hline
Decay channel & Candidates & \multicolumn{3}{c|}{Signal} \\
\hline
\hline
$\decskp$   &  5022~~~  & 4434 &  71&  55 \\
$\decskpp$  &  7327~~~  & 5429 &  86& 124 \\
$\deckp$    &  7682~~~  & 5032 &  97&  75 \\
$\decskppp$ & 14565~~~  & 8710 & 121& 276 \\
$\decskpS$  & 10766~~~  & 5824 & 104& 322 \\
$\dpkpp$    & 12664~~~  & 6357 & 113& 102 \\
\hline
\end{tabular}
\end{center}
\label{tabsel}
\end{table}

All combinations of two and four tracks or two tracks and a
$\pi^0$, with total null charge, are considered as $\D^0$ candidates, and all
combinations of three tracks with total charge $+1$ are considered as $\D^+$ candidates. 
The $\pi^0$\/ candidates are selected from 
two-photon combinations having a $\chi^2$\/ probability of at least 5\%
for a mass-constrained kinematical fit~\cite{SELE}.
The invariant mass of the D candidates, with mass assignment according to
particle hypothesis, is required to be close to the $\D$ meson mass
within two times the invariant mass resolution.
The $\deckp+(\pi^0)$ channel is reconstructed as
 the $\deckp$\/ decays, except
that candidates are kept 
if the $\K^- \pi^+$\/ invariant mass is between $1.5~\GeVcc$\/ and
$1.7~\GeVcc$\/.

\begin{figure}
 \setlength{\unitlength}{1.0mm}
 \begin{center}
  \begin{picture}(170,170) 
  \put(0,0){\mbox{\psfig{figure=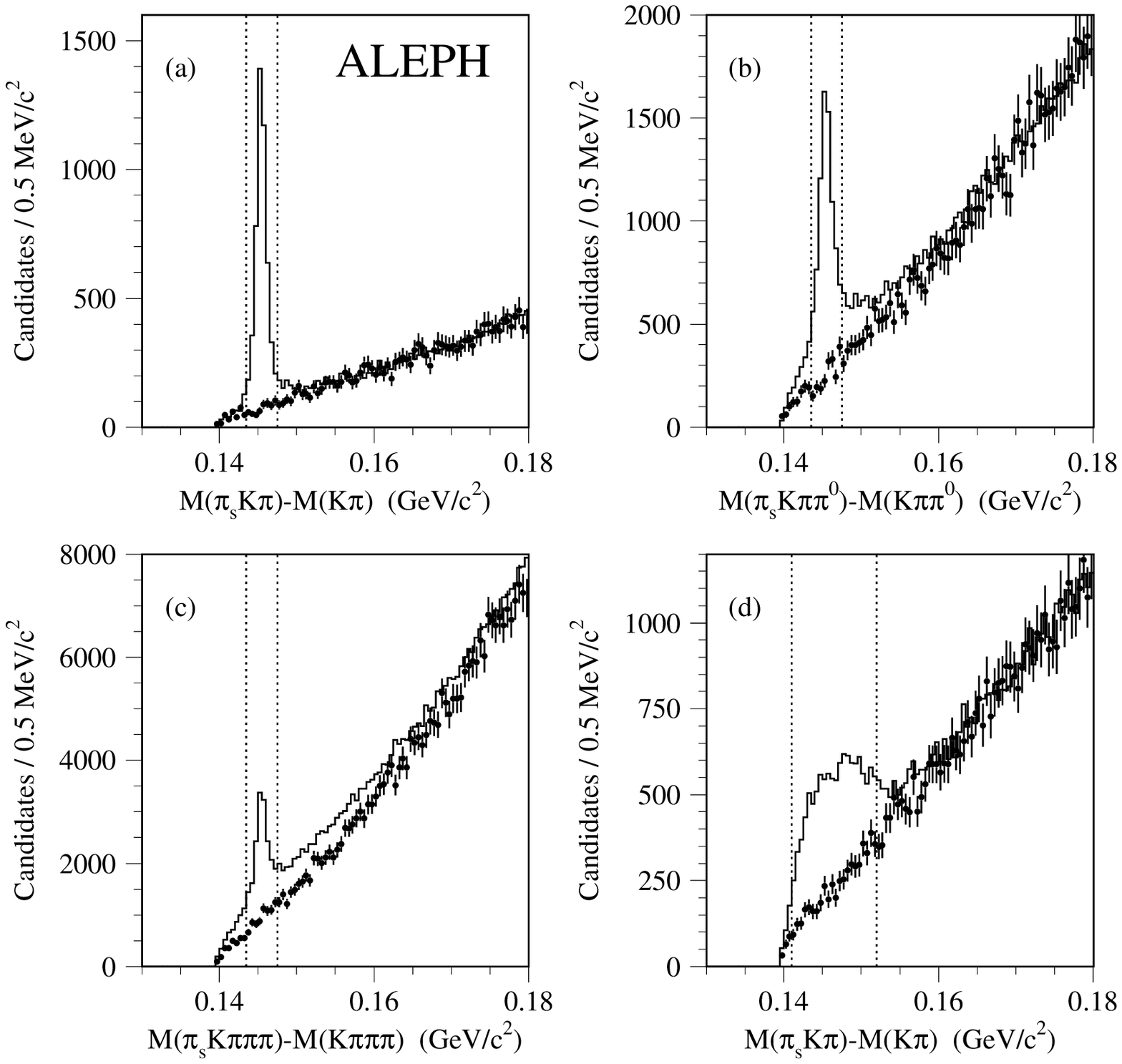,width=17cm,height=17cm}}}
  \end{picture}
 \end{center}
 \caption{\it  \label{figds}
   Mass-difference distribution for candidates of the decay channel 
   $\decdst$, followed by $\deckp$(a), $\deckpp$(b), $\deckppp$(c) and
   $\deckpS$(d) . The full histogram are the data while the dots with
   error bars are the combinatorial background taken from Monte Carlo
   simulation. The error bars represent both the statistical error due
   to the limited Monte Carlo statistics and the systematic error in
   the normalisation. The dotted lines show the selected region.} 
\end{figure}

$\D^{*+}$\/ candidates from  $\decdst$\/ decays are selected 
by adding an extra track with momentum less than $3.5~\GeVc$ to a $\D^0$
candidate. 
In the channels ({\it ii}), ({\it iv}) and ({\it v}) the combinatorial
background is reduced by requiring   
the $\D^0$ candidate to satisfy $| \cos \theta^* | \le 0.8$\/,
where $\theta^*$, in the $\D^0$\/ rest frame, is the angle between
the $\D^0$\/ direction and the 
sphericity axis of the three ($\deckpp$) or four ($\deckppp$) decay
products, or the kaon direction in the decay $\deckp(\pi^0)$ with an
undetected $\pi^0$.
The $\pi^+_s$ momentum is required to be greater than $1.5~\GeVc$,
favoring high momentum $\D^{*+}$'s, in
order to reject combinatorial background  
and $\Zbb$\/ events in which a b hadron decays into a $\D^{*+}$.
In Figure~\ref{figds} the mass difference $\Delta M = M_{\D^{*+}} -
M_{\D^0}$\/ distributions 
are shown. Candidates are selected in the 
$\Delta M$ region $143.5~\MeVcc$\/ to  $147.5~\MeVcc$\/ for the 
channels ({\it i}), ({\it ii}) and ({\it iv}) and 
$141~\MeVcc$\/ to $152~\MeVcc$\/ for channel ({\it v}).

Candidates from $\deckp$ decays, channel ({\it iii}), are
selected if the 
kaon track  momentum is greater than $2.5~\GeVc$ and the pion track momentum
greater than $1.5~\GeVc$. 
A common vertex is searched for and candidates are kept if a 
vertex with a $\chi^2$ probability greater than 1$\%$ is found, 
and the projected decay length significance is greater 
than unity. The asymmetry measurement in this channel suffers from
 the presence of a significant contribution 
of fake candidates due to incorrect mass assignments, which reverse the charge
 assignment of the charm quark. For a large fraction of these
 candidates, the correct mass assignment is also selected in the event. 
The $\dedx$ measurements of the two tracks are used to choose 
between the two mass combinations. The probability to be a kaon ($P_{\K}$)
 or a pion ($P_{\pi}$) is computed from the measured track ionization and 
the expectation for a kaon or a pion. The mass assignment which 
gives the highest probability $P_{\K} \times P_{\pi}$ is kept. If no
$\dedx$ measurement is available for the tracks, the choice is made
randomly. This criterion reduces the contribution from incorrect mass
assignments to $4\%$ in the signal region.  

$\D^{+}$ candidates, channel ({\it vi}), are selected if the kaon
track has a momentum greater  
than $2.5 ~\GeVc$ and if the $\dedx$ measurement is more  consistent with the  
expectation for a kaon than for a pion. One of the two pion tracks is required
to have a momentum greater than  $1.5 ~\GeVc$, 
the other pion momentum being greater than $0.75 ~\GeVc$.
The three tracks are required to form a common vertex with a $\chi^2$
probability greater than 1$\%$, and a significance of the decay length,
projected along the $\D^{+}$ momentum, greater than 1.5.
Finally, in case of multiple candidates, only the candidate with the largest
 decay length significance is kept.

\begin{figure}
 \setlength{\unitlength}{1.0mm}
 \begin{center}
  \begin{picture}(170,170)
  \put(0,0){\mbox{\psfig{figure=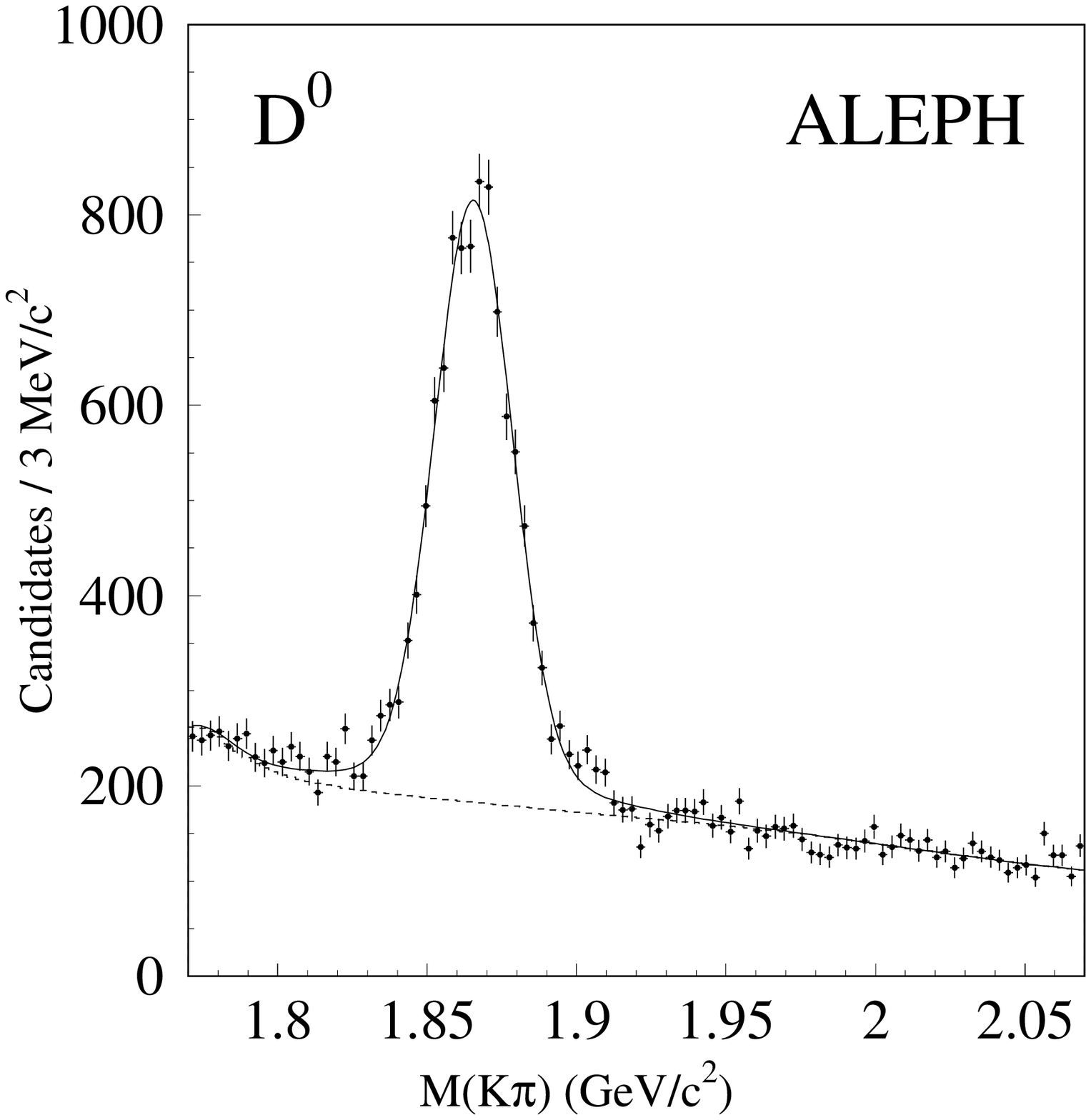,width=17cm,height=17cm}}}
  \end{picture}
 \end{center}
 \caption{\it \label{figd0}
   Mass distribution for candidates of the decay channel 
   $\deckp$. The dots with error bars are data while the solid line
   represent the best fit to the distribution and the dashed line is
   the best fit to the background.}
\end{figure}

\begin{figure}
 \setlength{\unitlength}{1.0mm}
 \begin{center}
  \begin{picture}(170,170)
  \put(0,0){\mbox{\psfig{figure=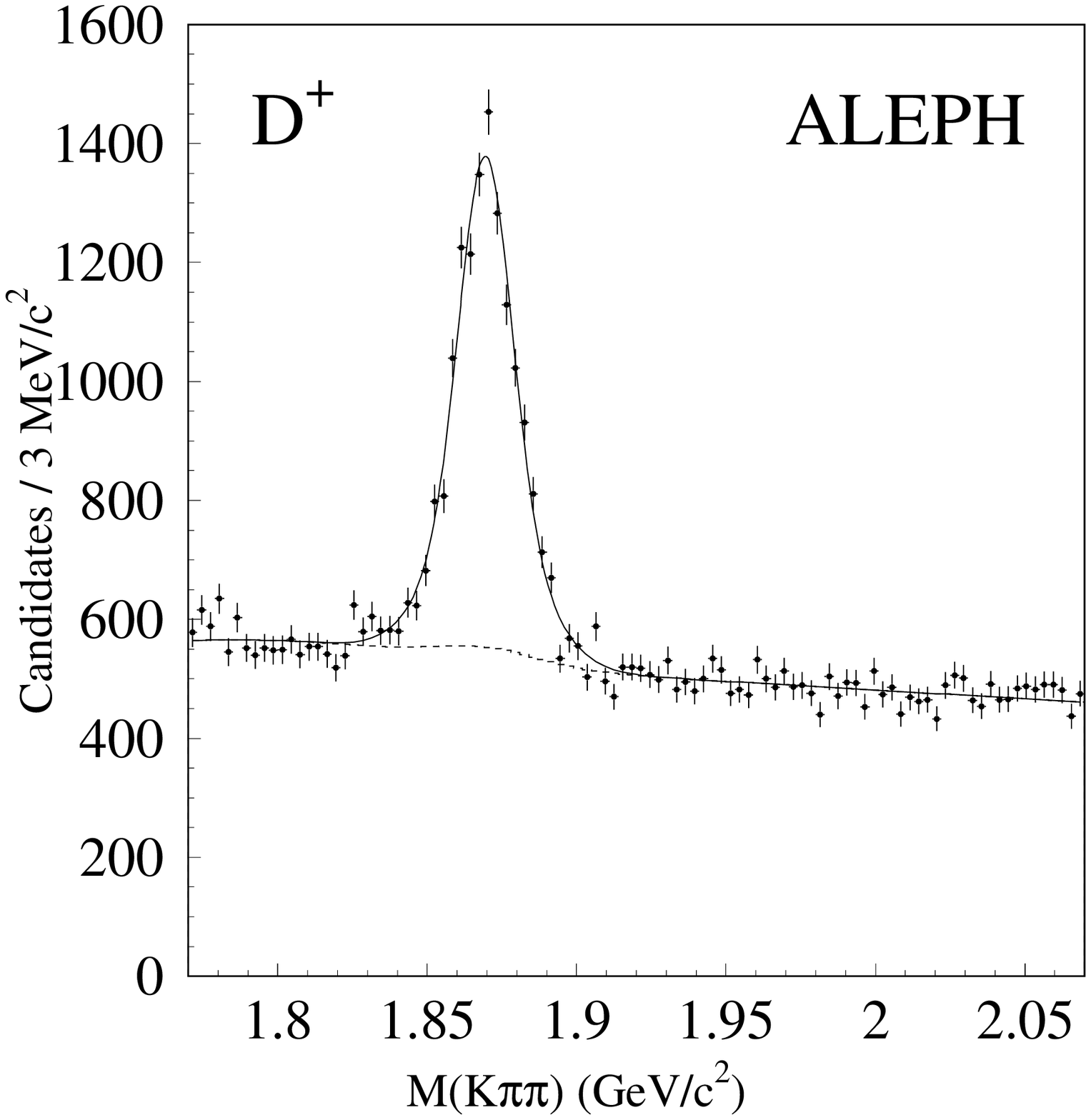,width=17cm,height=17cm}}}
  \end{picture}
 \end{center}
 \caption{ \label{figdp}\it
   Mass distribution for candidates of the decay channel 
   $\dpkpp$. The dots with error bars are data while the solid line
   represent the best fit to the distribution and the dashed line is
   the best fit to the background.}
\end{figure}

The contribution of the $\Zcc$\/ process to the $\deckp$ and $\dpkpp$ signals
 is enhanced to 80$\%$ by selecting $\D^0$ and $\D^+$ candidates with 
energy greater than half the beam energy. 
The resulting invariant mass distributions are shown in Figure~\ref{figd0}
for the $\deckp$ sample and in Figure~\ref{figdp} 
for the $\dpkpp$ sample.

\section{Combinatorial background estimate}

The fraction of combinatorial background events in the $\D^{*+}$
sample is estimated from the mass difference distributions.
The data sample contains, in addition to $\dstp$'s which are correctly
reconstructed, a combinatorial
background and a fraction of $\dstp$'s obtained from soft pions 
and partially reconstructed or fake $\D^0$'s. The latter contribution,
clearly seen in Figure~1(c), 
carries the correct charm quark charge, so it is treated as signal.
The mass difference of the combinatorial background is obtained from
$\D^0$ candidates in Monte Carlo simulated events in which no $\dstp$'s
have been produced. A track from fragmentation is added to such 
candidates and the combinatorial background is estimated from the resulting 
$\Delta M$ distribution, normalized to the data in the region $\Delta M >
0.16~\GeVcc$. In the background normalisation procedure reflections of
the signal in the $\Delta M > 0.16~\GeVcc$ region have to be taken
into account. This is done by using, both in data and in background
Monte Carlo, only events in which a signal candidate is not found.

The fraction of combinatorial background events in the $\D^{0}$ and
 $\D^{+}$ samples is extracted from a fit to the invariant 
 mass distributions (Figures~\ref{figd0} and \ref{figdp}).
The  $\D^{0}$ and  $\D^{+}$ signals are parametrized by
 two Gaussians with a common mean and the combinatorial background by a
polynomial function. Resonant background contributions, such as 
$\D^0 \to  \K^- \K^+, \pi^+ \pi^- (\pi^0)$ and $\deckp$ where 
the two mass assignments are reversed
($\D^0$ channel), $\D_{\s}^{\pm} \to \phi \pi^{\pm}$ and $\D_{\s}^{\pm}
\to  \K^* \K$ ($\D^+$ channel), are taken into account in the
fit. Their shapes and sizes  are fixed by the Monte Carlo simulation
with branching ratio according to PDG values~\cite{PDG}.

The fitted numbers of signal events in the different channels are shown in
Table~\ref{tabsel}.

\section{\mbox{Measurement of the forward-backward asymmetry}}

The measurement of the differential cross section for $\Zcc$ events
(Eq.~\ref{eq:dxsec}) requires the evaluation of 
the angle $\theta$ between the charm quark direction and the 
incident electron beam. This is measured from the
thrust axis, oriented along the candidate direction, $\cos\theta = -Q
\cos \theta_{\thr}$, $Q$ being the electric charge of the
reconstructed $\K$ in the $\D^{0}$ or $\D^{+}$ decays. 

Together with charm events two possible sources contribute 
to the observed asymmetry: combinatorial background and charm mesons 
from $\Zbb$ events. Therefore the observed asymmetry in the selected sample is
$$
\afb{\mathrm{obs}} = f_{\mathrm{sig}} f_{\c} \afb{\c} + f_{\mathrm{sig}}
(1-f_{\c}) \afb{\b} + (1-f_{\mathrm{sig}}) \afb{\mathrm{bkg}} \   
$$
where $f_{\mathrm{sig}}$ is the fraction of D mesons in the
sample, $f_{\c}$ is the fraction originating 
from direct charm production and $\afb{\mathrm{bkg}}$ is the
forward-backward asymmetry of the combinatorial background.
 
The fraction, $f_{\c}$, of D mesons originating from direct charm production is
measured directly from data~\cite{RC}. 
The event is divided into two hemispheres according to the thrust axis.
A lifetime-mass tag~\cite{RB} is applied on the hemisphere opposite to
the $\D$ meson, to select b hemispheres with 99$\%$ purity. 
The fraction, $f_{\mbox{\scriptsize b-tag}}$, of $\D$ mesons that survive 
the b-tag cut is used to extract
the charm fraction $f_{\c} = { {\displaystyle \epsilon_{\b\bar{\b}} -
 f_{\mbox{\scriptsize b-tag}}}
\over {\displaystyle \epsilon_{\b \bar{\b}} - \epsilon_{\c\bar{\c}}}}$, where
$\epsilon_{\b \bar{\b}}$ and $\epsilon_{\c\bar{\c}}$ are the b-tag efficiency for
b and charm events; $\epsilon_{\c\bar{\c}}$ is obtained from Monte
Carlo simulation, while $\epsilon_{\b \bar{\b}}$ is measured from an
unbiased $\Zbb$ data sample~\cite{RB} and corrected to take into
account the presence of an energetic $\D$ meson in the opposite
hemisphere. The systematic error on 
$f_{\c}$ arises mainly from the  uncertainty on this correction~\cite{RC}.
The measured charm fractions  are listed 
in Table~\ref{tabcfrac}.

\begin{table}[tb]
\caption[]{\it  Charm fraction and mixing probability used to extract
 the charm asymmetry from the observed asymmetry. The first
 column shows the charm fractions, as 
 measured from data, with the statistical and systematic
 uncertainties. The second column  shows the  obtained fractions of
 $\D^{* \pm}$'s, $\D^0$'s and $\D^+$'s coming from $B^0$  decays among
 all $\b \to  \D^{* \pm},\D^0,\D^+$ decays. The resulting values  of
 $\chi_{\mathrm mix}$ are listed in the last column. } 
\begin{center}
\begin{tabular}{|l|c|c|c|c|}
\hline
 \rule[-1.5ex]{0.mm}{2.em} Decay channel & $f_{\c}$ & ${{\mathrm{B}^0 \to \D}\over{\b \to \D}}$
& $\chi_{\mathrm mix}$ \\ 

\hline
\hline
$\decskp$   & $0.741\pm 0.019\pm 0.007$  & $0.80 \pm 0.05$ &  $0.16\pm 0.04$\\
$\decskpp$  & $0.743\pm 0.019\pm 0.007$  & $0.80 \pm 0.05$ &  $0.16\pm 0.04$\\
$\deckp$    & $0.787\pm 0.019\pm 0.006$  & $0.22\pm 0.05$ &  $0.035\pm 0.005$\\
$\decskppp$ & $0.783\pm 0.016\pm 0.006$  & $0.80 \pm 0.05$ &  $0.16\pm 0.04$\\
$\decskpS$  & $0.766\pm 0.021\pm 0.007$  & $0.80 \pm 0.05$ &  $0.16\pm 0.04$\\
$\dpkpp$    & $0.797\pm 0.020\pm 0.006$  & $0.72 \pm 0.03$ &  $0.12\pm 0.02$\\
\hline
\end{tabular}
\end{center}
\label{tabcfrac}
\end{table}

The b asymmetry $A_{FB}^{\b}$ 
is fixed to the Standard Model values at the three centre of mass
energies, as listed in Table~\ref{tabbasym}, together with the
dependence of the fitted charm asymmetry.
The effective asymmetry that enters in this analysis is diluted by a
factor $(1-2\chi_{\mathrm mix})$, 
due to the mixing of neutral b mesons. The mixing probability
$\chi_{\mathrm mix}$ is different for each reconstructed $\D$
meson, since it depends on the fraction of such  
mesons produced in  $\mathrm{B}^0$ decays among all $\b \to \D^{*\pm},
\D^0,\D^+$ decays. These fractions are derived from Monte Carlo
simulations and the associated  errors are taken 
as the difference between the 
Monte Carlo prediction and an estimate
based on experimental measurements in the semileptonic
sector~\cite{BDECAY}. These values, together with the  
world average value of $\chi_d$, are used to obtain the $\chi_{\mathrm mix}$
values, shown in Table~\ref{tabcfrac}.
The contribution from the double-charm decays of b
hadrons, in which the charge of the reconstructed $\D^{(*)}$ has the opposite
sign with respect to single-charm decays, is negligible due to
the low momentum of the decay products~\cite{BDD}.

\begin{table}[tb]
\caption{\it Value of $\afb{\b}$ at the three energy
  points, used to extract $\afb{\c}$. These value correspond to the
  Standard Model prediction with $m_t=175~\GeVcc$, $m_H=127~\GeVcc$\/
  and $\alpha_s=0.120$, without QCD correction in the final state (a
  discussion of the corrections to the asymmetry in this analysis
  follows in the text). The last column shows the dependence of the
  charm asymmetry on the value of $\afb{\b}$.} 
  $$
  \begin{array}{|l|c|c|}\hline
    \sqrt{s} (\GeV) & \afb{\b} (\%) & d\afb{\c}/d\afb{\b} \\ \hline\hline 
    89.37   &  5.7         &    -0.22          \\
    91.22   &  9.7         &    -0.22          \\
    92.96   &  12.1        &    -0.22          \\ \hline
  \end{array}
  $$
\label{tabbasym}
\end{table}

The doubly Cabibbo suppressed decay $\D^0\to\K^+\pi^-$, which affects
both the charm and the b component of the inclusive $\D^0$
sample, has a negligible effect on the results.
The asymmetries are corrected to take into account the
fraction  of the selected $\D^{(*)}$'s that originate from 
gluon splitting, estimated to be $(0.8\pm0.4)\%$~\cite{RC}. 

The asymmetry of the combinatorial background is measured 
from the upper side band of the mass peaks. 
Within the side bands, multiply-counted events
are selected only once by choosing the candidate randomly.
In the case of the $\D^+$, the contributions of the resonant backgrounds
 are negligible in the upper side band region. 
On the other hand, in the case of the $\deckp$ channel, 
the side band sample contains a contribution
 where the two mass assignments of the $\D^{0}$ decay products are reversed.
 This contribution induces an asymmetry opposite to the real asymmetry.
 Its size (around 5$\%$ of the side band sample)
 is estimated from the Monte Carlo simulation and then 
subtracted from the measured
 background asymmetry, using the Standard Model values for the charm and b
 asymmetries. The correction shifts the measured $\D^{0}$ background asymmetry by 0.004.
The combinatorial background asymmetries for all channels, measured at the 
three centre of mass energies, are listed in Table~\ref{tababkg}.

\begin{table}[tb]
\caption[]{\it Combinatorial background asymmetries, measured on data, at the 
three centre of mass energies, from the side band of the mass peaks. 
}
\begin{center}
\begin{tabular}{|l|r@{$\pm$}r|r@{$\pm$}r|r@{$\pm$}r|}
\hline
Decay channel & \multicolumn{2}{c|}{peak $-$ 2 $\GeV$}  & \multicolumn{2}{c|}{peak}
& \multicolumn{2}{c|}{peak + 2 $\GeV$} \\ 
\hline
\hline
$\decskp$   & $ 0.001 $&$ 0.045$ & $ 0.0050 $&$ 0.0092 $&$ -0.013 $&$ 0.036$\\
$\decskpp$  & $ 0.016 $&$ 0.024$ & $ 0.0063 $&$ 0.0050 $&$  0.045 $&$ 0.019$\\
$\deckp$    & $-0.105 $&$ 0.060$ & $-0.005  $&$ 0.013  $&$ -0.015 $&$ 0.048$\\
$\decskppp$ & $ 0.021 $&$ 0.014$ & $ 0.0035 $&$ 0.0029 $&$  0.011 $&$ 0.011$\\
$\decskpS$  & $ 0.036 $&$ 0.028$ & $-0.0026 $&$ 0.0057 $&$  0.006 $&$ 0.021$\\
$\dpkpp$    & $ 0.016 $&$ 0.024$ & $ 0.0015 $&$ 0.0052 $&$ -0.004 $&$ 0.020$\\

\hline
\end{tabular}
\end{center}
\label{tababkg}
\end{table}

The charm asymmetry
is extracted by means of an unbinned maximum likelihood
fit giving the following results at the three different centre of mass
energies:  
$$
\begin{array}{lrcl}
A_{FB}^{\c}(\sqrt{s} = 89.37~\GeV)&=&\left( -1.0\pm 4.3 \pm 1.0\right)\!\%  \\
A_{FB}^{\c}(\sqrt{s} = 91.22~\GeV)&=&\left(  6.3\pm 0.9 \pm 0.3\right)\!\%  \\
A_{FB}^{\c}(\sqrt{s} = 92.96~\GeV)&=&\left( 11.0\pm 3.3 \pm
  0.8\right)\!\% \ .  
\end{array}
$$
 
\begin{figure}
 \setlength{\unitlength}{1.0mm}
 \begin{center}
  \begin{picture}(170,170)
  \put(0,0){\mbox{\psfig{figure=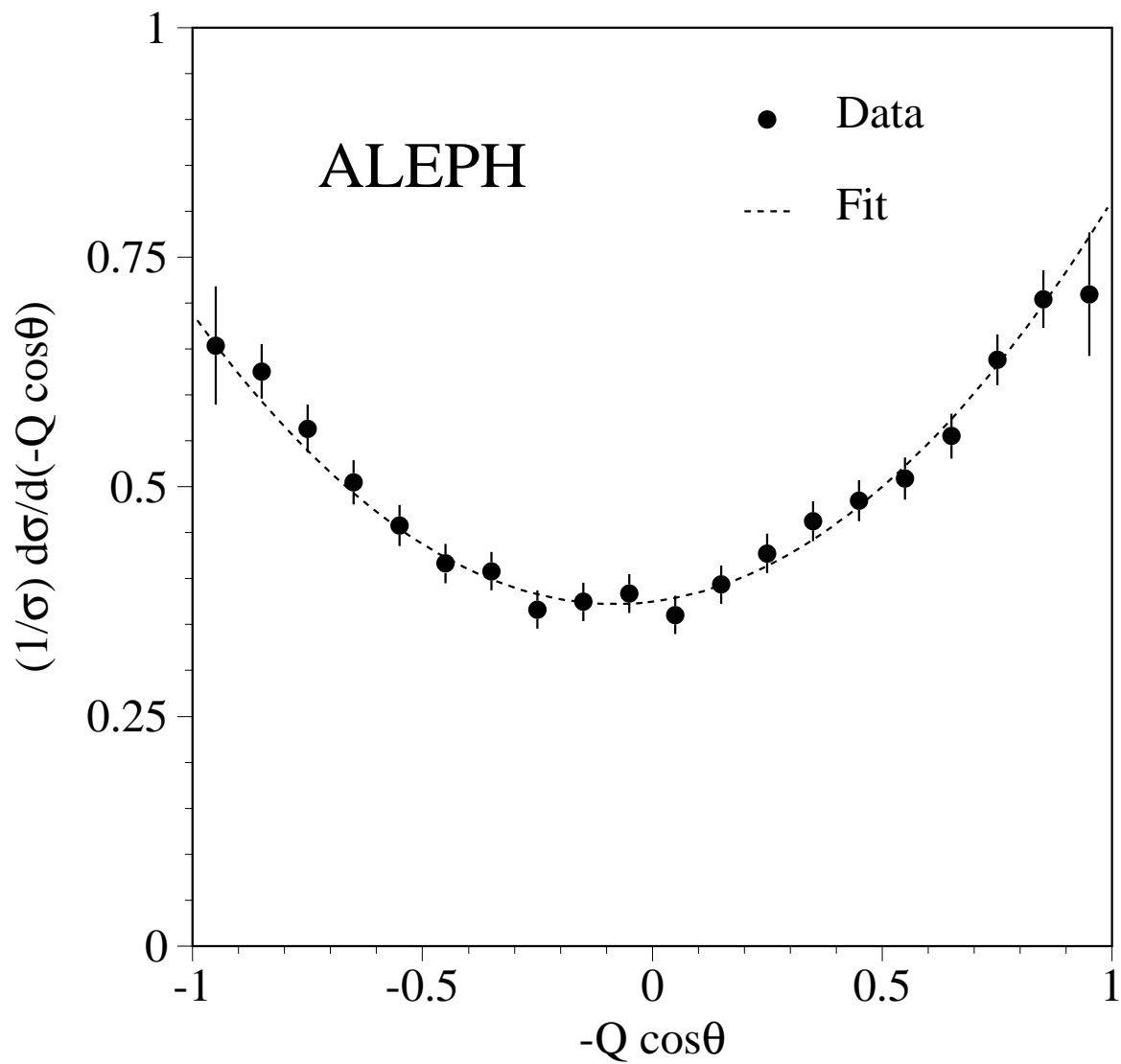,width=17cm,height=17cm}}}
  \end{picture}
 \end{center}
 \caption{ \label{figxsec}\it
  Normalized angular distribution of the tagged $\Zcc$ events at
  $\sqrt{s} = 91.2~\GeV$, corrected
  for acceptance and background subtracted. The dots with error bars are data 
   while the dashed line is the result of the fit.}
\end{figure}

\begin{figure}
 \setlength{\unitlength}{1.0mm}
 \begin{center}
  \begin{picture}(170,170)
  \put(0,0){\mbox{\psfig{figure=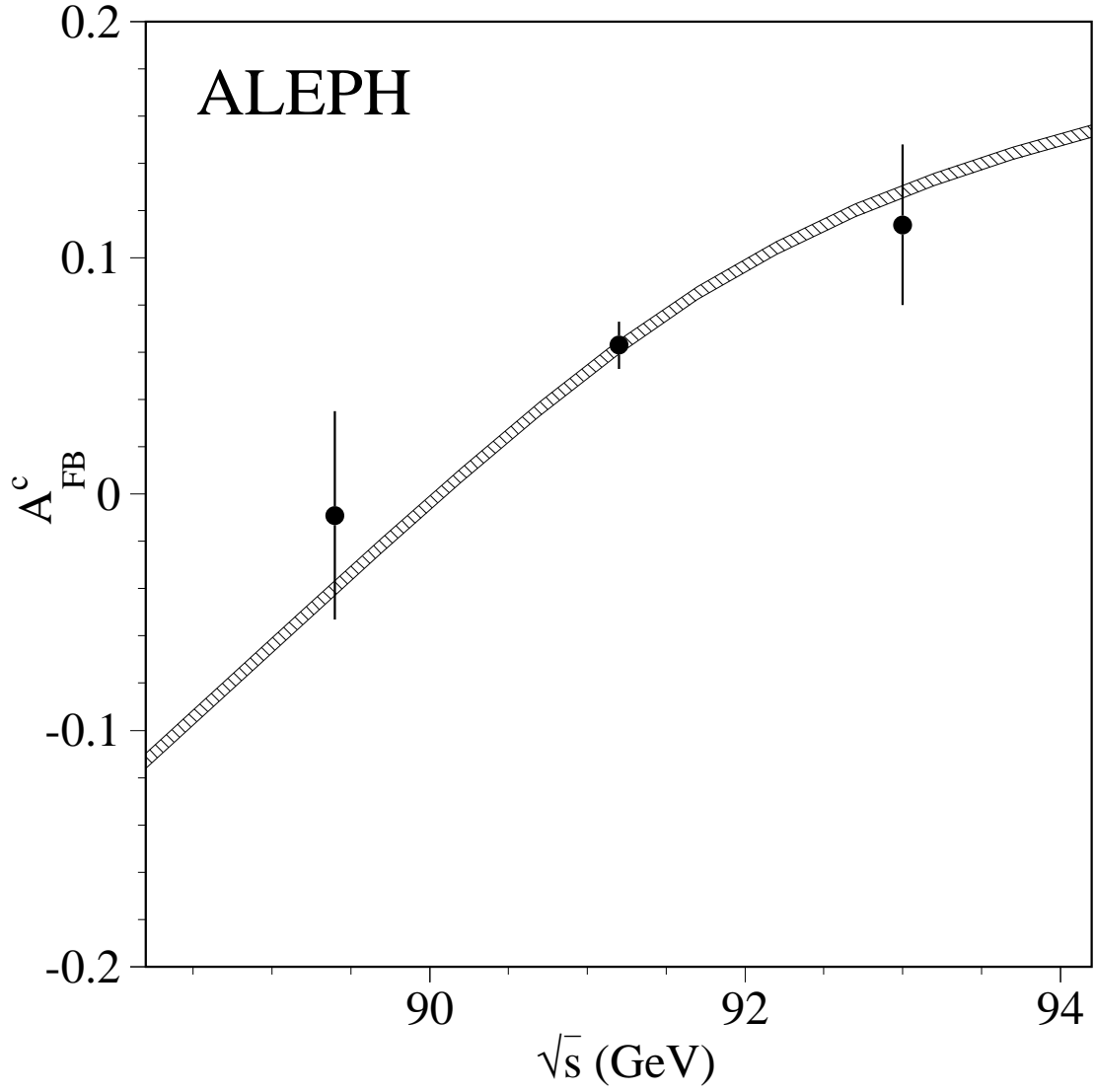,width=17cm,height=17cm}}}
  \end{picture}
 \end{center}
 \label{figasym}
 \caption{\it $A_{FB}^{\c}$ as a function of the 
centre of mass energy. The dots with error bars are the 
measured asymmetries.
The curve is the prediction of the Standard Model
with $m_t=175~\GeVcc$, $m_H=[90,1000]~\GeVcc$\/ and $\alpha_s=0.120$.}
\end{figure}

The first error is statistical and the second arises from systematic
uncertainties as listed in Table~\ref{tabsyst}.
The angular distribution of the tagged $\Zcc$ events at the Z
is shown in Figure~\ref{figxsec}, after background subtraction
and acceptance corrections. 
Figure~\ref{figasym} shows the measured asymmetries as a function of the
centre of mass energy  together with the predictions 
of the Standard Model.

\begin{table}[t]
\caption[]{\it Sources of systematic errors on the measured charm
  forward-backward  asymmetry. The total errors are obtained summing
  in quadrature the relative contributions.}
\begin{center}
\vspace{1em}
\begin{tabular}{|l|c|c|c|}
\hline
{ Source}& \multicolumn{3}{|c|}{ $\Delta \afb{\c}\ (\%)$} \\
 & { peak $-$ 2 $\GeV$} & { peak} & { peak + 2 $\GeV$}  \\
\hline\hline
{ Fraction of D mesons $f_{\mathrm{sig}}$}  &$\pm$ 0.15 &$\pm$ 0.10
&$\pm$ 0.22 \\ 
{ Charm fraction $f_{\c}$}  &$\pm$ 0.10 &$\pm$ 0.02  &$\pm$ 0.12 \\
{ b mixing      }  &$\pm$ 0.09 &$\pm$ 0.17  &$\pm$ 0.18 \\
{ Comb. back. asymmetry}  &$\pm$ 0.98 &$\pm$ 0.20  &$\pm$ 0.75 \\
{ Gluon splitting} &    ---    &$\pm$ 0.03  &$\pm$ 0.05 \\
\hline\hline
{ TOTAL } &$\pm$ 1.00 &$\pm$ 0.28 &$\pm$ 0.81 \\\hline
\end{tabular}

\end{center}
\label{tabsyst}
\end{table}

The pole asymmetry as defined in Eq.~2 is extracted from the
measured asymmetries at the three energy points by expressing them as a 
single measurement at the Z mass and applying a correction for the
effect of initial and final state radiation, QCD corrections and
photon exchange and interference. 
As pointed out in Ref.~\cite{QCD}, the theoretical estimate of the QCD correction
has to be rescaled to take into account the bias from the experimental
cuts. In particular, the requirement of high momentum $\D^{(*)}$s
removes events in which hard gluon emission occurred, substantially
reducing the correction.  
As computed from Monte Carlo, the relative QCD correction after selection cuts 
are found to be consistent with zero within a $0.14\%$ absolute
uncertainty. The 
initial state radiation correction can also be biased by the selection
cuts, which are less efficient at lower centre of mass energies. This
correction, calculated to be 14.9\% using the MIZA program~\cite{miza},
is lowered to ($14.1\pm0.1$)\% when the selection bias is taken into
account. The final relative correction to be applied to the measured asymmetry 
at the energy of the Z mass is $(13.2\pm0.2)\%$. 

Within the Standard Model, the measured asymmetry can be used to
extract a value of $\sin^2\theta_W^{\mathrm eff}$, taking into account
the dependence of the b asymmetry on this quantity, yielding
$$
\sin^2 \theta_W^{\mathrm eff} = 0.2321 \pm 0.0016 \ . 
$$

\section{Conclusion}
The forward-backward asymmetry in $\Zcc$ decays has been measured at three different energies,
at the Z peak and off-peak at $\pm$ 2 GeV:
$$
\begin{array}{rcl}
A_{FB}^{\c}(\sqrt{s} = 89.37~\GeV)&=&\left( -1.0\pm 4.4\right)\!\%  \\
A_{FB}^{\c}(\sqrt{s} = 91.22~\GeV)&=&\left( 6.3\pm 1.0\right)\!\%  \\
A_{FB}^{\c}(\sqrt{s} = 92.96~\GeV)&=&\left( 11.0\pm 3.4\right)\!\% \ . 
\end{array}
$$

A Standard Model fit to the measured asymmetries yields 
$\sin^2 \theta^{\mathrm eff}_W = 0.2321 \pm 0.0016$. These 
measurements are in agreement with the Standard Model predictions and
with the other determinations of these parameters at LEP~\cite{leprb2}.

\section*{Acknowledgements}

We thank our colleagues from the accelerator divisions for the
successful operation of the \LEP\ machine and the engineers 
and technical staff in all our institutions for their contribution
to the good performance of \A . Those of us from non-member
states thank {\sc CERN} for its hospitality.

\pagebreak

\pagebreak

\pagebreak

\pagebreak

\pagebreak

\end{document}